\documentclass[american,letterpaper,conference,compsoc]{IEEEtran}
\pdfoutput=1
\usepackage[T1]{fontenc}
\usepackage[latin9]{inputenc}
\usepackage{array}
\usepackage{prettyref}
\usepackage{booktabs}
\usepackage{multirow}
\usepackage{amstext}
\usepackage{graphicx}

\makeatletter

\providecommand{\tabularnewline}{\\}

\usepackage{microtype}

\IEEEoverridecommandlockouts

\usepackage[shortcuts]{extdash}

\newrefformat{fig}{Fig.~\ref{#1}}

\def\IEEEbibitemsep{.9pt plus .5pt}

\let\origsection\section
\renewcommand\section{\@ifstar{\starsection}{\nostarsection}}
\newcommand\nostarsection[1]
{\sectionprelude\origsection{#1}\sectionpostlude}
\newcommand\starsection[1]
{\sectionprelude\origsection*{#1}\sectionpostlude}
\let\origsubsubsection\subsubsection
\renewcommand\subsubsection{\@ifstar{\starsubsubsection}{\nostarsubsubsection}}
\newcommand\nostarsubsubsection[1]
{\sectionprelude\origsubsubsection{#1}\sectionpostlude}
\newcommand\starsubsubsection[1]
{\sectionprelude\origsubsubsection*{#1}\sectionpostlude}
\newcommand\sectionprelude{%
  \vspace{-.9mm}
}
\newcommand\sectionpostlude{%
  \vspace{-.9mm}
}

\@ifundefined{showcaptionsetup}{}{%
 \PassOptionsToPackage{caption=false}{subfig}}
\usepackage{subfig}
\makeatother

\usepackage{babel}
\begin{document}

\title{DETOx: Towards Optimal Software-based Soft-Error Detector Configurations
\thanks{This work was supported by the German Research Council (DFG) focus
program SPP~1500 under grant SP~968/5-3.}}

\author{\IEEEauthorblockN{Michael Lenz and Horst Schirmeier}
\IEEEauthorblockA{Department of Computer Science 12\\
Technische Universität Dortmund, Germany\\
e-mail: \{michael.lenz, horst.schirmeier\}@tu-dortmund.de}}
\maketitle
\begin{abstract}
Application developers often place executable assertions -- equipped
with program-specific predicates -- in their system, targeting programming
errors. However, these detectors can detect data errors resulting
from transient hardware faults in main memory as well. But while an
assertion reduces silent data corruptions (SDCs) in the program state
they check, they add runtime to the target program that increases
the attack surface for the remaining state. This article outlines
an approach to find an optimal subset of assertions that minimizes
the SDC count, without the need to run fault-injection experiments
for every possible assertion subset.
\end{abstract}

\section{Introduction}

With continuously shrinking semiconductor structure sizes and lower
supply voltages, the per-device susceptibility to transient hardware
faults is on the rise. A class of countermeasures with growing popularity
is \emph{software-implemented} hardware fault tolerance (SIHFT), which
avoids expensive hardware mechanisms and can be applied application\-/specifically.
However, SIHFT can, against intuition, cause more harm than good:
its overhead in execution time and memory space also increases the
system's figurative ``attack surface''. In consequence, this phenomenon
can diminish all gains from detected or corrected errors by the increased
possibility of being struck by radiation in the first place \cite{schirmeier:15:dsn}.

One class of SIHFT measures are \emph{executable assertions} capable
of detecting data errors. These statements or code sequences check
statically-known state invariants in specifically chosen points of
a workload at runtime, and \emph{``can detect errors in input data
and prevent error propagation''}~\cite{saib:1977:executable-assertions}.
Although primarily used as a means to complement unit tests -- aiming
at detecting programming errors during the development process especially
of safety-critical software, -- assertions can also detect data errors
caused by hardware faults, or even be specifically designed for this
purpose.

While, for example, Hiller et al. \cite{hiller:2002:placement-executable-assertions}
identify good placements and adequate predicates for executable assertions,
we assume the target workload is already equipped with a set of assertions.
We explore the tradeoff between each assertion's capability to detect
data errors, and its runtime cost that increases the liveness of critical
data in variables residing in memory during their execution. We also
assume that data errors may be detected by more than one assertion,
and aim at finding a subset of assertions -- an assertion \emph{configuration}
-- that catches most errors but minimizes the total \emph{silent data
corruption} (SDC) count.

\section{The Attack-Surface Tradeoff}

\begin{figure}
\begin{centering}
\subfloat[\label{fig:faultspace_both}Assertions~1 and~2 both compiled in.]{\begin{centering}
\includegraphics[width=6cm]{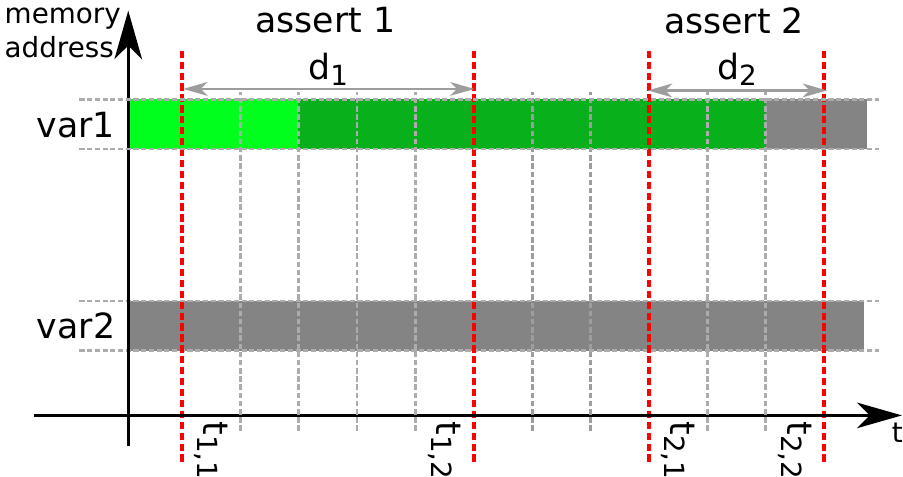}
\par\end{centering}

}
\par\end{centering}

\begin{centering}
\vspace{-2mm}
\subfloat[\label{fig:faultspace_2only}Only assertion~2 compiled in: The runtime
reduces by $d_{1}$, the total gray area is reduced compared to \prettyref{fig:faultspace_both}.]{\begin{centering}
\includegraphics[width=6cm]{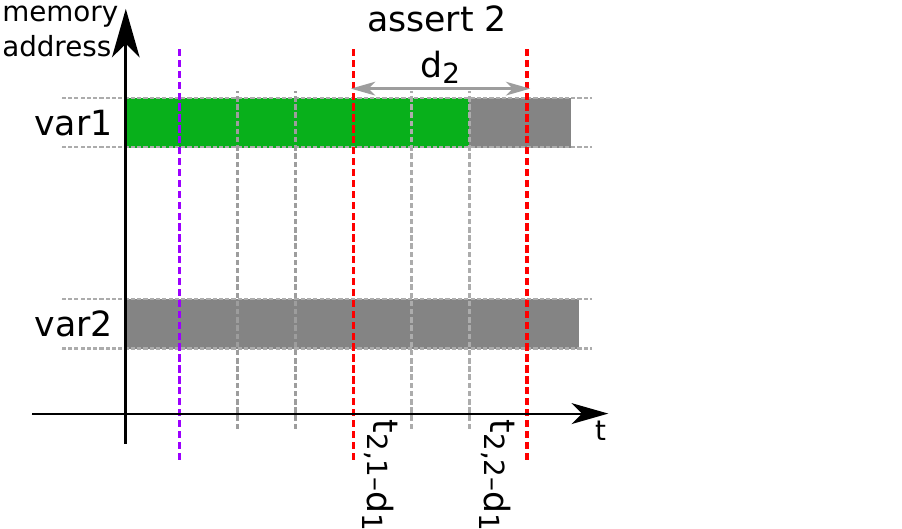}
\par\end{centering}

}\caption{\label{fig:faultspaces}Fault space of an example workload in two
assertion configurations: Gray areas denote SDCs, dark green areas
show faults detected by assertion~2, and light green indicates detection
by both assertions.}

\par\end{centering}

\end{figure}

We assume a fault model of uniformly distributed, independent and
transient single-bit flips in main memory, modeled as originating
from \emph{direct} influences of ionizing radiation \cite{sridharan:2013:fengshui-s}.
Under this assumption, live data stored in a variable in memory has
a corruption probability that is proportional to its lifetime during
the workload run and its memory size \cite{schirmeier:15:dsn} --
or, in other words, proportional to its \emph{area} in a fault-space
diagram.

The X-axis of the example diagram in \prettyref{fig:faultspace_both}
shows the workload's runtime, the Y-axis all bits in memory. In the
example, \emph{var2} holds live data throughout the complete runtime,
and a fault-injection (FI) campaign injecting faults at all possible
time\slash{}memory-bit locations yields the information that bit-flips
in this variable during this time frame lead to SDCs (color-coded
\emph{gray}). The gray-colored area contributes to the total SDC count
of the workload.

\subsubsection*{Error Detection and Additional Runtime}

Similarly, \emph{var1} holds live data, but is checked against an
application-specific predicate by executable assertions~1 and~2
at runtime. Hence, corruptions occurring in \emph{var1} before the
point in time when assertion~2 runs are \emph{detected} (color-coded
\emph{green}). However, both assertions entail a runtime cost: The
instructions executed for each assertion extend the workload runtime
(by $d_{1}$ respectively $d_{2}$). Thereby, these assertions prevent
corruptions in \emph{var1} from causing an SDC, but increase the
lifetime of the data stored in \emph{var2} -- which in turn enlarges
the grey-colored area, or the total SDC count.

In consequence, depending on each assertion's balance between SDC
\emph{reduction} (by turning otherwise gray \emph{SDC} areas into
green \emph{detected} results) and SDC \emph{increase} (by increasing
the lifetime of other variables), it  may influence the workload's
total SDC count positively or negatively. However, in the example
in \prettyref{fig:faultspace_both}, assertion~1 detects a subset
of the data errors in \emph{var1} that assertion~2 detects as well.
Thus, it seems advisable to generate a workload variant \emph{without}
assertion~1, yielding a different fault-space diagram (\prettyref{fig:faultspace_2only})
with a lower total runtime and, hence, a lower (\emph{gray}) SDC count
originating from \emph{var2}, but still most data errors in \emph{var1}
being detected by assertion~2.

Unfortunately, assertions cannot be considered independently: Depending
on the assertion \emph{configuration} -- i.e., other assertions being
compiled in or out of a specific workload variant -- an assertion
may increase the fault-space area of a protected (yielding green \emph{detected}
results) or an unprotected variable (gray \emph{SDC} results). This
interdependency makes it necessary to consider the total SDC count
resulting from \emph{every possible} assertion configuration.

\subsubsection*{Approach: FI-based Result Calculation}

Considering real-world workloads with the number of $N$ different
executable assertions in the hundreds or thousands makes clear, that
generating a workload variant for each of the $2^{N}$ possible assertion
configurations is generally infeasible. Running an FI campaign measuring
the total SDC count for each of these workload variants even exponentiates
this problem.

Instead, our approach is based on FI results from a single workload
variant with \emph{all} assertions enabled. We exhaustively scan the
single-bit flip fault space of this workload (using advanced pruning
methods within the FAIL{*} FI tool \cite{schirmeier:15:edcc-s}) and
record a result -- for example, \emph{No Effect} or \emph{SDC} --
for each fault-space coordinate. If an FI experiment observes that
one of the assertions detected a data error, we do not abort the experiment
and record \emph{detected} (\emph{green} in \prettyref{fig:faultspaces}),
but record \emph{which} assertion triggered \emph{and let the experiment
continue running}. So, for example, an FI experiment injecting in
\emph{var1} at some point in time before assertion~1 runs (\prettyref{fig:faultspace_both})
would record assertion~1 to have detected the error, continue running,
then also record assertion~2, and finally record that an SDC occurred
-- because it \emph{would} have occurred if none of the two assertions
had been in place.

Based on this result data, our DETOx tool prototype can \emph{calculate}
the SDC count for an arbitrary assertion configuration. Removing,
e.g., assertion~2 from the example in \prettyref{fig:faultspace_both}
requires 1)~\emph{subtracting} all gray SDC areas between $t_{2,1}$
and $t_{2,2}$ from the total SDC count, and 2)~\emph{``re-dyeing''}
all remaining areas that were only detected by assertion~2 to the
final result recorded in the FI experiments.

\section{Sorting Example}

Using our approach we evaluated the configurations of a simple sorting
program taking 24 input elements with two assertions: the first one
checks for ascending order of two swapped elements, while the second
repeats the same test on the complete array after sorting. The following
table shows the FI-obtained SDC count for the all-enabled (``11'')
variant, and the predicted results for all other configurations. Assertion
configurations are represented as a bit vector, where the bits indicate
whether either assertion was enabled/disabled.\smallskip{}

\noindent%
\begin{tabular}{rcccc}
\toprule 
 & \textbf{\small{}00} & \textbf{\small{}01} & \textbf{\small{}10} & \textbf{\small{}11}\tabularnewline
\midrule
{\small{}$\textrm{{SDC}}_{\textrm{{prediction}}}$} & \textbf{\small{}2\,315\,851} & \textbf{\small{}1\,395\,495} & \textbf{\small{}2\,318\,461} & \multirow{2}{*}{{\small{}1\,547\,549}}\tabularnewline
\cmidrule{1-4} 
{\small{}$\textrm{{SDC}}_{\textrm{{reality}}}$} & {\small{}2\,319\,929} & {\small{}1\,393\,233} & {\small{}2\,324\,435} & \tabularnewline
\midrule
{\small{}error} & {\small{}-0.176\,\%} & {\small{}+0.162\,\%} & {\small{}-0.257\,\%} & \tabularnewline
\bottomrule
\end{tabular}\smallskip{}

To quantify the prediction quality, we ran FI campaigns for all other
configurations besides ``11'' as a ground truth for comparison --
information that would usually not be available, shown in the {\small{}$\textrm{{SDC}}_{\textrm{{reality}}}$
row. }In this example, the SDC-count predictions are accurate to within
$0.3$\,\%. The optimal, lowest SDC-count configuration is ``01'',
i.e., only the second assertion gets enabled.

\section{Conclusions}

To conclude, our approach allows for fast and cheap exploration of
the assertion-configuration space, and is based on FI results of a
single, all-enabled configuration. This allows searching for optimal
configurations in future work, for example using genetic algorithms
or optimization techniques such as integer-linear programming.

\bibliographystyle{IEEEtran}
\bibliography{bib/macros-short,bib/all,local}

\end{document}